# $^4$He glass phase: a model for liquid elements

Robert F. Tournier[1,2] and Jacques Bossy[1,2]

[1]Univ. Grenoble Alpes, Inst. NEEL, F-38042 Grenoble Cedex 9, France
[2]CNRS, Inst. NEEL, F-38042 Grenoble Cedex 9, France
E-mail address: robert.tournier@creta.cnrs.fr

**PACS: 65.60.+a; 64.70.pm; 67.25.bd**

**Abstract:** The specific heat of liquid helium confined under pressure in nanoporous material and the formation, in these conditions, of a glass phase accompanied by latent heat are known. These properties are in good agreement with a recent model predicting, in liquid elements, the formation of ultrastable glass having universal thermodynamic properties. The third law of thermodynamics involves that the specific heat decreases at low temperatures and consequently the effective transition temperature of the glass increases up to the temperature where the frozen enthalpy becomes equal to the predicted value. The glass residual entropy is about 23.6% of the melting entropy.

## Introduction

The solid-liquid transformation of bulk helium depends on the pressure p and temperature T [1-6]. The melting entropy is determined from the Clapeyron relation knowing the volume and melting temperature changes associated with the solid-to-liquid transformation under a pressure p [2]. The phase diagram is deeply modified when liquid helium is confined in 25Å mean diameter nano-porous media under pressures p where $3.58 \leq p \leq 5.27$ MPa [7]. Early studies of these involved specific heat anomaly measurements and they were viewed as a consequence of the formation of localized Bose-Einstein condensates on nanometer length scales analogous to a solid [7]. The glass phase has since been discovered using measurements of the static structure factor, S(Q), of helium confined in the porous medium MCM-41 with pore diameter 47±1.5 Å. A similar amorphous S(Q) was also observed in 34 Å Gelsil [8]. The presence of an amorphous phase has been confirmed at higher pressures using porous Vycor glass [9]. In this work we consider that the supercooled liquid far below the melting temperature $T_m$ is condensed in a glass phase accompanied by an exothermic latent heat associated with a first-order transition.

Recent work reviews earlier findings of glass formation in pure metals of small size and thickness [10-21]. It is clear that there is a need for a fundamental understanding of the resistance to crystallization of these glasses [21]. A recent model predicting the thermodynamic properties of the ultra-stable vitreous phase of liquid elements is used for that purpose [22]. Liquid helium at high pressures is a normal liquid having properties analogous to those of liquid elements with much higher melting temperatures. Nevertheless, its zero point enthalpy $H_0$ is far from negligible at low temperatures [1, 23]. The influence of the enthalpy change $\Delta H_0$ has to be evaluated as a function of pressure to confirm that vitreous helium can be compared to other glasses. In addition the proximity to absolute zero and the third law of thermodynamics reduces the values of the specific heat of helium below and around the glass transition temperature $T_g$. Such influence does not occur in other liquids. A recent conference has been devoted to glass and entropy [24] pointing out a theoretical need to clarify whether residual entropy exists in the glass state at low temperatures [25]. In this work the model previously used for liquid elements having higher melting temperatures [22] is applied to these problems.

**The model and its application to $^4$He under pressure**

A complementary negative contribution $-v \times \Delta p = -v \times \varepsilon_{ls} \times \Delta H_m / V_m$ depending linearly on $\theta^2 = (T-T_m)^2 / T^2_m$, has been added to the classical Gibbs free energy change $\Delta G_{1ls}$ for liquid-to-solid transformation. $\Delta H_m$ is the melting enthalpy per mole, $\varepsilon_{ls}$ a fraction of $\Delta H_m$, $V_m$ the molar volume, $\Delta p$ the complementary Laplace pressure and $v = 4\pi/3 \times R^3$ the solid nucleus volume of radius R. The new $\Delta G_{ls}$ per mole is given by (1):

$$\Delta G_{ls} = 4\pi R^3 3^{-1} \Delta H_m \times (\theta - \varepsilon_{ls}) + 4\pi R^2 (1 + \varepsilon_{ls}) \sigma_{1ls} \Delta H_m. \tag{1}$$

$\Delta G_{ls}$ is associated with solid nucleus formation in a melt; $\varepsilon_{ls}$ is the critical enthalpy saving coefficient, $(1+\varepsilon_{ls}) \times \sigma_{1ls}$ the new surface energy and $\sigma_{1ls}$ the classical surface energy for $\varepsilon_{ls}$ =0 [26]. This complementary enthalpy explains the presence of intrinsic long-lived metastable nuclei surviving above $T_m$ and disappearing as the applied superheating rate increases [27, 28]. Crystallization and melting are initiated by the formation of these solid or liquid growth nuclei accompanied by a volume change that is expected to obey Lindemann's rule in pure liquid elements [29]. Lindemann's description shows that the ratio of the mean square root $\Delta R$ of thermal vibrations and the interatomic distance R is a universal constant $\delta_{ls}$ at the melting temperature $T_m$.

The complement $\varepsilon_{ls0} \times \Delta H_m$ associated with the growth nucleus formation at the melting temperature $T_m$ ($\theta$ = 0) has been determined for many pure liquid elements and glass-forming melts. The coefficient $\varepsilon_{ls0} = 0.217$ is the same for many pure liquid elements [30] but is much larger than 1 and smaller than 2 in many glass-forming melts [26]. Lindemann's constant $\delta_{ls} = 0.103$ is directly deduced from $\varepsilon_{ls0} = 0.217$ when the classical Gibbs free energy change is extended to include this new contribution [22].

The glass state has been described as a thermodynamic equilibrium between crystal and liquid states [26]. The glass transition is induced by an enthalpy change. The critical enthalpy saving coefficients $\varepsilon_{ls}(\theta)$ and $\varepsilon_{gs}(\theta)$ for the formation of a solid growth nucleus at the reduced temperature $\theta$ in the liquid state and in the glass state are given by (2) and (3) where $\theta_{0m}$ and $\theta_{0g}$ are the reduced temperatures at which the enthalpy saving becomes zero and $T_K$ is the Kauzmann temperature:

$$\varepsilon_{ls}(\theta) = \varepsilon_{ls0}(1 - \theta^2 \times \theta_{0m}^{-2}). \tag{2}$$

$$\varepsilon_{gs}(\theta) = \varepsilon_{gs0}(1 - \theta^2 \times \theta_{0g}^{-2}) \qquad \text{for } T_K \leq T \leq T_g. \tag{3}$$

The enthalpy change per mole below $T_g$ from the glass to the liquid is $[\varepsilon_{ls}(\theta) - \varepsilon_{gs}(\theta)] \times \Delta H_m$. The frozen enthalpy below $T_g$ has a maximum value equal to $(\varepsilon_{ls0} - \varepsilon_{gs0}) \times \Delta H_m$ in strong glasses. Following cooling below $T_g$ it takes centuries for the glass to become fully relaxed at the Kauzmann temperature, (point at which the frozen enthalpy is greatest), despite an easy recovery at $T_g$ during heating. Therefore it is time-consuming to attain the thermodynamic equilibrium from the glass phase by relaxation. Various microscopic models prove the existence of a phase transition at $T_g$ [31-40]. Here, we only use thermodynamic relations without considering the microscopic aspects of the liquid-glass transformation.

The universal value $\delta_{ls} = 0.103$ of the Lindemann constant obtained for pure metallic elements at their melting temperatures $T_m$ has been used to build a model for their vitrification [22]. Its extension to liquid helium is proposed. The Gibbs free energy change below $T_g$ cannot include any variation in structural relaxation enthalpy because $\delta_{ls}$ cannot be lowered below its minimum value. The enthalpy difference per mole is characterized by $\theta_{0m} = -2/3$ ($T_{0m} = T_m/3$), $\theta_{og} = -1$ ($T_{0g} = 0$), $\varepsilon_{ls0} = \varepsilon_{gs0} = 0.217$, and is proportional to $\Delta\varepsilon_{lg}$ in (4) above $T_m/3$ and in (5) below $T_m/3$:

$$\Delta\varepsilon_{lg} \times \Delta H_m = -\left[0.217 \times 1.25 \times \theta^2\right]\Delta H_m. \tag{4}$$

$$\Delta\varepsilon_{lg} \times \Delta H_m = -0.36165 \times (1+\theta) \times \Delta H_m \tag{5}$$

The specific heat jump is maximum at $T = T_m/3$ and constant below $T_m/3$ because the contribution given by (2) becomes equal to zero below this temperature.

The glass transition temperature in liquid elements is $0.3777 \times T_m$ when the frozen enthalpy and the latent heat can be accommodated by the glass and liquid enthalpy [22]. In liquid helium, because of the specific heat reduction near the absolute zero, this is not possible. It has been shown that an enthalpy excess $\Delta\varepsilon \times \Delta H_m$ obtained after rapid quenching or vapor deposition increases $\varepsilon_{gs}$ ($\theta$) in the glass state [41] as shown in (6):

$$\varepsilon_{gs}(\theta) = \varepsilon_{gs0}(1 - \theta^2 \times \theta_{0g}^{-2}) + \Delta\varepsilon. \tag{6}$$

The reduced temperature $\theta_2 = [\varepsilon_{gs}(\theta_2)-2]/3$ of homogeneous nucleation [29] which is equal to the glass transition $\theta_{g2}$ [26] is now a solution of the new quadratic equation (7):

$$\theta_{g2}^2 \times \varepsilon_{gs0} \times \theta_{0g}^{-2} + 3\theta_{g2} + 2 - \varepsilon_{gs0} - \Delta\varepsilon = 0. \tag{7}$$

The value of $\theta_2 = \theta_{g2}$ is given by (8) [40]:

$$\theta_{g2} = (-3 + \left[9 - 4(2 - \varepsilon_{gs0} - \Delta\varepsilon)\varepsilon_{gs0} / \theta_{0g}^2\right]^{1/2})\theta_{0g}^2 / (2\varepsilon_{gs0}). \tag{8}$$

The enthalpy excess coefficient $\Delta\varepsilon$ also exists in liquid helium-4 because of the specific heat reduction due to the third law of thermodynamics. In liquid elements described by (8) and $\Delta\varepsilon = 0$, we have $\theta_{g2} = \theta_g = -0.6224$ or $T_g = 0.3777 \times T_m$.

In spite of the specific heat reduction, the frozen enthalpy in the glass phase has to be equal to $0.105 \times \Delta H_m$ even if $T_g$ is larger than $0.3777 \times T_m$. If the available glass enthalpy cannot attain $0.105 \times \Delta H_m$ at $T_g$, an excess enthalpy corresponding to $\Delta\varepsilon = 0.105$ would exist above this temperature during cooling. A new glass transition temperature given by (8) has to take place at $\theta_{g2}$ = −0.58398. The new latent heat would be equal to 0.0925×$\Delta H_m$ at this temperature instead of 0.105×$\Delta H_m$ as given by (4). In fact, the effective glass transition still occurs above $\theta_{g2}$ up to the temperature $T_{geff}$ where the frozen enthalpy becomes equal to 0.105×$\Delta H_m$. The endothermic latent heat 0.0925×$\Delta H_m$ begins to be recovered at the same temperature.

The specific heats of confined liquid helium at pressures of 3.58, 4.45, 4.89 and 5.27 MPa have been measured by [7] and are shown in Figure 1:

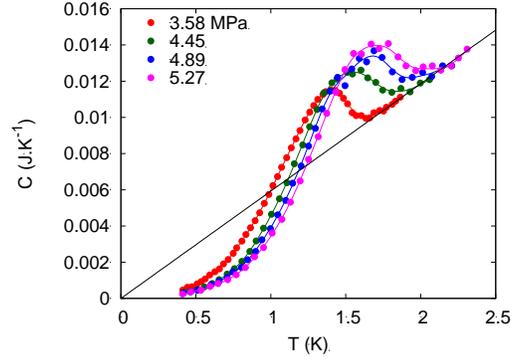

**Figure 1:** Fig. 2 of Yamamoto et al [7]. Copyright 2016 by the American Physical Society. Specific heat of confined helium-4 versus temperature at various pressures and added line $\gamma T$ =0.00593×T.

The liquid specific heat in the cell does not have a strong dependence on pressure and is approximately proportional to the temperature T, so we use the approximation $\gamma T$ with $\gamma$ =0.00593. This introduces some error in the difference $\Delta C_p = C_{pl}$-$C_{pg}$ of liquid and glass specific heats as shown in Figure 2 around 2 K.

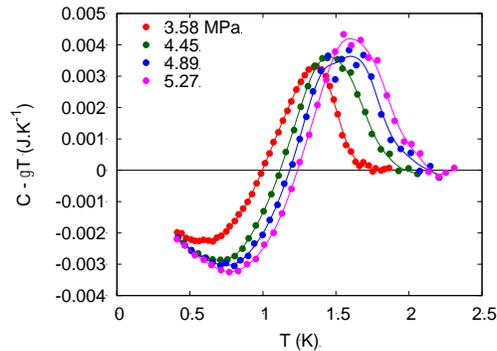

**Figure 2:** Specific heat difference (C-$\gamma T$) of confined helium-4 versus temperature T (K) with $\gamma$=0.00593×T. $H_2$×V/$V_m$ the area below zero and $H_1$×V/$V_m$ the area above zero.

The volume of confined helium in nano-pores is calculated using the linear specific heat measured at 2.28 MPa and is represented in Figure 1a and 1c of [7] knowing that the volume open to bulk helium-4 is 38.5 mm³. This gives V = 60.3 mm³. The molar volume $V_m$ under pressure is interpolated from known values [3] using (9):

$$V_m \ (cm^3.mole^{-1}) = -2.246 \times lnp + 23.176. \tag{9}$$

where p is the pressure in MPa.

The molar specific heat of liquid is calculated as a function of the mole fraction contained in the volume V and is equal to $\gamma T * V_m/V$. The values of $\gamma$ and $V_m/V$ are given in Table 1.

## Table 1: Application of the model to $^4$He under pressure

1- The pressure p in MPa, 2- $V_m/V$ the molar volume divided by the confined helium volume, 3- $T_m$ the melting temperature of helium crystals, 4- $H_2$ the frozen enthalpy in the glass phase 5- $H_1$ the enthalpy of fusion of the glass phase 6- $\Delta V_m$ the volume change at $T_m$, 7- $V_m$ the molar volume of the liquid, see (9), 8- $S_m$ the total entropy change at $T_m$, see (15), 9- $\Delta S_m$ the melting entropy, see (12), 10- $S_{RL}$ the residual entropy at zero K, see (17), 11- $s_R$, see (16), 12- $S_{Rg}$ the residual entropy of the glass phase, 13- $\gamma T$ the liquid specific heat as observed in Figure 1, 14- $\theta_D$ the Debye temperature of solid helium, see (11), 15- $T_{g2}$ the calculated glass transition temperature, see (8) with $\Delta\varepsilon = 0.105$, 16- $T_{m0}$ the onset temperature of glass melting [7], 17- $\Delta H_m$ the melting enthalpy equal to $\Delta S_m \times T_m$, 18- $H_2/\Delta H_m$, 19- $H_1/\Delta H_m$, 20- $T_{geff}$ the theoretical temperature where the glass enthalpy becomes equal to that of the liquid, 21- $\Delta H_0$ the enthalpy at zero K, 22- $\Delta U_0$ the internal energy at zero K.

| 1 | p (MPa) | 3.58 | 4.45 | 4.89 | 5.27 |
|---|---|---|---|---|---|
| 2 | $V_m/V$ | 359.3 | 350.3 | 346.6 | 343.5 |
| 3 | $T_m$ (K) | 1.937 | 2.157 | 2.268 | 2.364 |
| 4 | $H_2$ (J.mole$^{-1}$) | 0.52 | 0.696 | 0.766 | 0.843 |
| 5 | $H_1$ (J.mole$^{-1}$) | 0.449 | 0.574 | 0.654 | 0.717 |
| 6 | $\Delta V_m$(cm$^3$) | 1.353 | 1.3 | 1.29 | 1.27 |
| 7 | $V_m$(cm$^3$) (liq) | 21.66 | 21.12 | 20.9 | 20.71 |
| 8 | $T_K$ (K) | 0.497 | 0.599 | 0.624 | 0.658 |
| 9 | $\gamma$(J.K$^{-2}$.mole$^{-1}$) | 2.13 | 2.077 | 2.055 | 2.037 |
| 10 | $\Delta S_m$ (J.K$^{-1}$.mole$^{-1}$) | 2.75 | 2.87 | 2.99 | 3.06 |
| 11 | $S_m$ (J.K$^{-1}$.mole$^{-1}$) | 4.72 | 4.99 | 5.13 | 5.25 |
| 12 | $s_R$ (J.K$^{-1}$.mole$^{-1}$) | 0.908 | 0.878 | 0.859 | 0.845 |
| 13 | $S_{RL}$ (J.K$^{-1}$.mole$^{-1}$) | 1.968 | 2.121 | 2.141 | 2.185 |
| 14 | $S_{Rg}$ (J.K$^{-1}$.mole$^{-1}$) | 1.06 | 1.23 | 1.21 | 1.24 |
| 15 | $\theta_D$ (K) | 24.67 | 26.1 | 26.83 | 27.46 |
| 16 | $T_{g2}$ (K) calc | 0.806 | 0.897 | 0.944 | 0.983 |
| 17 | $T_{m0}$ (K) exp | 0.86 | 0.93 | 0.92 | 0.988 |
| 18 | $\Delta H_m$(J.mole$^{-1}$) | 5.33 | 6.19 | 6.77 | 7.23 |
| 19 | $H_2/\Delta H_m$ | 0.098 | 0.112 | 0.113 | 0.116 |
| 20 | $H_1/\Delta H_m$ | 0.084 | 0.093 | 0.097 | 0.099 |
| 21 | $T_{geff}$ (K) | 1.164 | 1.378 | 1.436 | 1.511 |
| 22 | $\Delta H_0$ (J.mole$^{-1}$) | 5.60 | 6.53 | 7.01 | 7.44 |
| 23 | $\Delta U_0$ (J.mole$^{-1}$) | 0.759 | 0.743 | 0.704 | 0.751 |

The solid specific heat $C_{vS}$ at constant volume depends on the Debye temperature according to (10) and the difference between $C_{pS}$ and $C_{vS}$ is considered to be negligible:

$$C_{vS} = 1952 \times T^3 \times \theta_D^{-3}.$$ (10)

The Debye temperature $\theta_D$ is interpolated from known values [3] using (11):

$$\theta_D(K) = 1.649 \times p(MPa) + 18.76.$$ (11)

The difference $C_p$-$C_v$ of liquid specific heats becomes smaller and smaller as the temperature decreases and at T = 2.5 K is only 3% of $C_v$ [6]. In addition, the compressibility of the liquid and solid are nearly equal when the solid is in equilibrium with the liquid [1]. At this point the specific heat values $C_V$ are considered as being equal to $C_p$ at low temperatures.

The specific heat minimum in Figure 2 occurs at T = $T_m/3$ = 0.646, 0.719, 0.756, and 0.788 K as predicted. The experimental minimum values: 0.810, 1.008, 1.051 and 1.107 are equal to or smaller than the theoretical ones: 0.995, 1.038, 1.081 and 1.107 J/K/mole. A consequence of the third law is that, instead of being constant and equal to 0.36165×$\Delta S_m$ [22], the specific heat $\Delta C_p(T)$ falls sharply below $T_m/3$. The temperatures $T_{g2}$ deduced from (7), using the enthalpy excess coefficient $\Delta \varepsilon$ =0.105, are given in Table 1 and are close to the observed temperatures $T_{m0}$ for the onset of glass phase melting [7] as shown in Table 1 and Figure 3. The calculated temperatures $T_{geff}$ are also closed to the freezing onset temperatures $T_{FO}$ [42, 43].

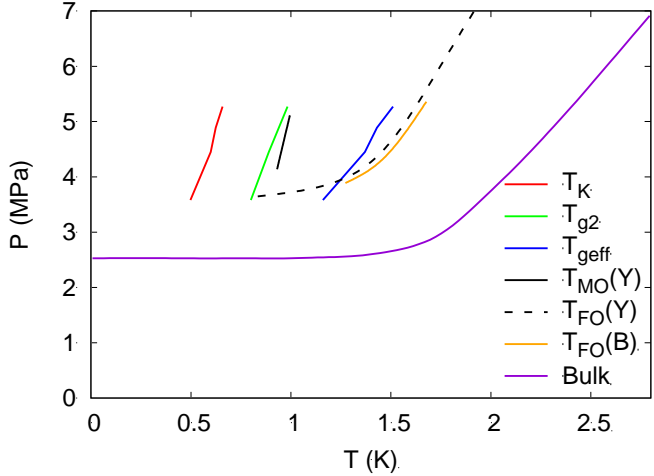

**Figure 3**: The P-T phase diagram of $^4$He in 2.5-nm and 2.4-nm nanoporous glasses determined in our present study and in [42, 43]. The Kauzmann temperature is represented by the line $T_K$. The melting onset temperature $T_{MO}$ is compared to the glass transition temperature $T_{g2}$ and the freezing onset temperature $T_{FO}$ to the effective glass transition temperature $T_{geff}$. The temperatures $T_{FO}$ (Y), $T_{MO}$ (Y) and $T_{FO}$ (B) have been measured by Yamamoto et al and Bittner et al respectively.

The frozen enthalpy $H_2$ is determined from the area below the zero line and above the curves in Figure 2 using the multiplicative coefficients $V_m/V$ given in Table 1 together with the melting temperature $T_m$, and the melting heat $\Delta H_m$. The ratios $H_2/\Delta H_m$ are given in Table 1. Their mean value 0.11 is in good agreement with the predicted value of 0.105. The endothermic latent heat is still determined by the area below the zero line and below the specific heat curves in Figure 2. The ratios $H_1/\Delta H_m$ for the melting enthalpy are given in Table 1. Their mean value of 0.0932 is in good agreement with the theoretical value of 0.0925. Other measurements indicate hysteresis between the melting temperature window observed by heating and that of freezing observed by cooling [42, 43], confirming the first-order character of the glass transition. The radius reduction of nano-pores decreases the freezing onset temperature [43]. This phenomenon is due to an increase of the Laplace pressure reducing the applied pressure p in the nanopores, the melting temperature $T_m$ and the glass transition temperature $T_{g2}$ in quantitative agreement with the surface tension of $^4$He.

The entropy $\Delta S$ deduced from the specific heat in Figure 1 is represented as a function of temperature in Figure 4 for the four pressures used.

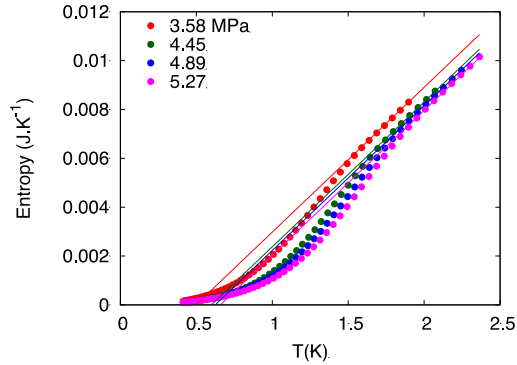

**Figure 4:** The entropy $\Delta S$ of confined helium-4 versus the temperature T (K) for four pressures. The Kauzmann temperature $T_K$ is extrapolated with the slope 0.00593 from the upper points.

The Kauzmann temperatures $T_K$ are deduced by extrapolation of the straight lines in Figure 4 down to zero entropy assuming that the solid entropy is negligible even if the experimental glass entropy is not fully negligible below these temperatures. The values of $T_K$ are given in Table 1. The melting enthalpy $\Delta H_m$ is obtained using (12) and given in Table 1:

$$\Delta S_m = \gamma(T_m - T_K) - S_S \tag{12}$$

where $S_S$ is the solid entropy at $T_m$.

So our first conclusion is that the enthalpy excess of $0.105 \times \Delta H_m$ (equal to the frozen enthalpy) and the endothermic latent heat of $0.0925 \times \Delta H_m$, characterize the thermodynamic parameters at the glass transition temperature $T_{g2}$. These correspond to the experimental values $T_{m0}$ given in Table 1 of the glass phase melting onset [7]. The endothermic latent heat is predicted to be fully recovered at the temperature $T_{geff}$ given in Table 1 where the frozen enthalpy attains its theoretical value of $0.105 \times \Delta H_m$ as shown in Figure 5 for p =3.58 (line 2) and 5.27 MPa (line 3):

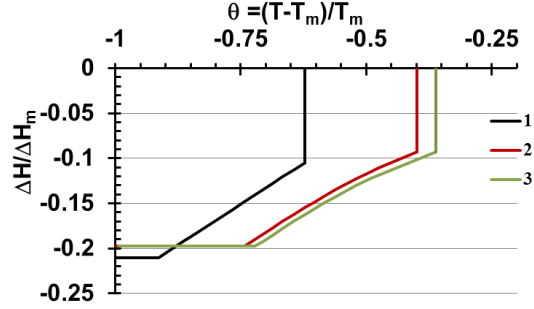

**Figure 5**: Outline of the theoretical enthalpy ratio $\Delta H/\Delta H_m$ versus $\theta = (T-T_m)/T_m$. Line 2 for p =3.58 MPa, $T_{geff}$ = 1.164 K, $T_K$ = 0.497 K and Line 3 for p =5.27 MPa, $T_{geff}$ =1.51 K, $T_K$ =0.658 K. Line 1 for other liquid elements.

All the enthalpy values represented in Figure 5 include the frozen enthalpy and the endothermic latent heat associated with the ultra-stable glass-to-liquid transformation. The increase of $\Delta H$ is determined using (13) from the Kauzmann temperature $\theta_K$ to $\theta = -2/3$ and (14) for $\theta \geq -2/3$:

$$\Delta H/\Delta H_m = 0.36165 \times (\theta - \theta_K) - 0.1975, \qquad (13)$$

$$\Delta H/\Delta H_m = 0.217 \times 1.25 \times (\theta^2 - 4/9) + A, \qquad (14)$$

where A is a constant equal to $\Delta H/\Delta H_m$ given by (13) for $\theta = -2/3$. Equations (13) and (14) are deduced from (4) and (5) including the frozen enthalpy and the latent heat. The enthalpy of ultra-stable glasses in other liquid elements are also represented in Figure 5 line 1 [22] with demonstrating an exception of the $3^{rd}$ law of thermodynamics.

**Residual entropy and enthalpy at 0 K**

Up to now we have used only the entropy $\Delta S$ and the enthalpy associated with the specific heat. The residual entropy and enthalpy at 0 K do not modify these as their derivatives are equal to zero. The amplitude of atomic vibrations at 0 K is large and cannot be neglected in liquid helium. Lindemann's rule introduces supplementary mean square amplitude of thermal vibrations to allow the melting of crystals. This phenomenon is associated with specific heat. The residual entropy and enthalpy at 0 K have a strong influence on the melting entropy $S_m$ and on the melting enthalpy $H_m$. The question of the existence of residual entropy is an important theoretical problem in considering the nature of the glass phase [24]. The properties of vitrified helium help clarify it.

Knowing $\Delta V_m$ and dP/dT, the solid-liquid melting entropy differences $S_m = (S_L - S_S)$ in $J.K^{-1}.mole^{-1}$ have been measured or calculated from the Clapeyron relation [2, 4, 5]. They obey the relationship given in (15) in the range 1.772K to 2.5 K, where the melting entropy $S_m$ and the melting temperature $T_m$ are as given in Table 1.

$$S_m = S_L - S_S = 1.235 \times T_m + 2.33. \qquad (15)$$

where $S_S$ is the crystal entropy at $T_m$.

The liquid entropy is approximated by (16)

$$S_L = \gamma T + s_R,$$  (16)

as shown in Figure 6 for p =3.58 and 5.27 MPa. $s_R$ is as given in Table 1.

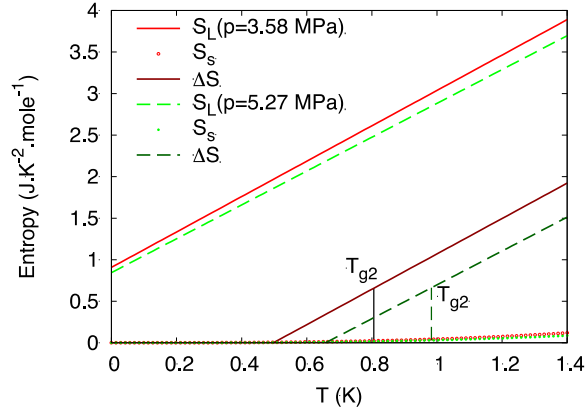

**Figure 6:** Outline of liquid helium-4 entropy $S_L$ and $\Delta S$ versus temperature T. The upper straight lines represent the total entropy $S_L$ (T) of liquids, including the residual entropy. The lower entropy lines $\Delta S$ (T) are deduced from the specific heat measurements. The broken lines correspond to p =5.27 MPa and the continuous lines to p =3.58 MPa. The points represent the solid entropies $S_S$ (T). The entropy changes $\Delta S$ at $T_{g2}$ are also indicated.

The liquid entropy $S_L$ is given by (17) as a function of the Kauzmann temperature $T_K$ and of its residual entropy $S_{RL}$:

$$S_L = \gamma(T-T_K) + S_{RL}.$$  (17)

$S_{RL}$ is equal to $\gamma T_K + s_R$ and to about 42% of $S_m$. The reduction of $S_{RL}$ by the glass formation entropy $(0.0925+0.105) \times \Delta H_m / T_{geff}$ at 0 K leads to the residual entropy $S_{Rg}$ of the glass given in Table 1. This experimental result demonstrates the existence at 0 K of residual entropy associated with structural disorder in glass equal to 23.6% of $S_m$ and 56% of the residual entropy of the liquid at zero K [24, 25]. The third law of thermodynamics predicts that the entropy change associated with a reversible transformation towards a thermodynamic equilibrium approaches zero as the temperature tends to zero. Residual entropy still remains in spite of the thermodynamic transition existence because the glass appears as an intermediate phase between liquid and crystal phases.

The total enthalpy change $H_m$ during the transformation from solid to liquid at $T_m$ is given by (18)

$$H_m = \gamma T_m^2 \times 2^{-1} + \Delta H_0 - H_S = T_m(S_L - S_S),$$  (18)

where $\Delta H_0$ is the difference of enthalpy at 0 K between the liquid and solid under the same pressure. $\Delta H_0$ is due to the volume change $\Delta V_m$. Values of $\Delta H_0$ are given in Table 1 for the four pressures used. The mean ratio 0.56 J.cm$^{-3}$ of $\Delta U_0/\Delta V_m$ is obtained subtracting the contribution p×$\Delta V_m$ in agreement with $\Delta U_0/\Delta V_m$ =0.5 J.cm$^{-3}$ of solid helium under pressure in the same range of molar volumes [1]. There is no enthalpy contribution associated with $\varepsilon_{ls}$×$\Delta H_m$ tending to zero when the atom number n of superclusters increases [26].

The entropy change at $T_{g2}$ in Figure 6 is still too small to accommodate the endothermic latent heat and the frozen enthalpy. However it could be achieved at $T_{geff}$ after a long time of isothermal relaxation. For example, the transformation at $T_g$ of an ultra-stable glass of indomethacin in supercooled liquid has been observed after several hours of isothermal relaxation [44].

**Conclusions**

We have shown that the vitreous transition in liquid helium has a thermodynamic origin and is accompanied by a latent heat. There is no enthalpy relaxing below the glass transition down to the Kauzmann temperature because this enthalpy is delivered at the glass transition temperature and leads to a first-order transition. This helium glass is then ultra-stable. Our model of thermodynamic transition predicts the values of the latent heat, the frozen enthalpy, the specific heat minimum of $\Delta C_p$ (T) at T = $T_m$/3, and the glass transition temperature of liquid elements.

This strongly suggests that the crystal growth nucleus formation is accompanied by a complementary enthalpy saving and that the classical nucleation equation has to be completed to be valid. In addition, we have confirmed that the glass entropy contains residual entropy associated with residual structural disorder at very low temperatures.

**Acknowledgments**: Our publication is based on specific heat measurements of K. Yamamoto, Y. Shibayama, and K. Shirahama [7]. The authors thank Prof. K. Shirahama from Keio University in Yokohama authorizing the reproduction of Figure 2.

# References

1. D.O. Edwards and R.C. Pandorf, *Phys. Rev.* **140** (1965) A816.

2. E.R. Grilly, *J. Low Temp. Phys.* **11** (1973) 33.

3. A. Driessen, E. van der Poll and I.F. Silvera, *Phys. Rev. B.* **33** (1986) 3269.

4. W.H. Keesom and P.H. Keesom, *Physica* **2** (1935) 557.

5. E.R. Grilly and R.L. Mills, *Ann. Phys.* 1**8** (1962) 250.

6. R.D. Mc Carty, *J. Phys. Chem.* **2** (1973) 923.

7. K. Yamamoto, Y. Shibayama and K. Shirahama, *Phys. Rev. Lett.* **100**, (2008) 195301.

8. J. Bossy, T. Hansen and H.R. Glyde, *Phys. Rev. B.* **81** (2010) 184507.

9. S. Bera, J. Maloney, N. Mulders, Z.G. Cheng, M.H.W. Chan, C.A. Burns and Z. Zhang, *Phys. Rev. B.* **88** (2013) 054512.

10. L. Zhong, H. Wang, H. Sheng, Z. Zhang and S. X. Mao, *Nature* **512** (2014) 177.

11. W. Klement Jr, R.H. Willens and P. Duwez, *Nature* **187** (1960) 869.

12. H.A. Davies and J.B. Hull, *J. Mater.Sci.* **11** (1976) 215.

13. W. Buckel and R. Hisch, *Z. Phys.* **138** (1954) 109.

14. K.H. Berhrndt, *J. Vac. Sci. Technol.* **7** (1970) 385.

15. D.M. Galenko and D.M. Herlach, *Mat. Sci. Eng. A.* ***34*** (2007) 449.

16. S.R. Corriel and D. Turnbull, *Acta Metall.* **30** (1982) 2135.

17. M.J. Aziz and W.J. Boettinger, *Acta Metall. Mater.* **42** (1994) 527.

18. J.Q. Broughton, G.H. Gilmer and K.A. Jackson, *Phys. Rev. Lett.* **49** (1982) 1496.

19. Y.W. Kim, H.M. Lin and T.F. Kelly, *Acta Metall. Mater.* **37** (1989) 247.

20. Y.W. Kim and T.F. Kelly, *Acta Metall. Mater.* **39** (1991) 3237.

21. A.L. Greer, *Nature Mater.* **14** (2015) 542.

22. R.F. Tournier, *Chem. Phys. Lett.* **651** (2016) 198.

23. A. Landesman II, *J. Phys. France, Colloq. C3.* **10** (1970) 55.

24. L. Wondraczek and R. Conradt, (Eds), Proceedings of the International Workshop on Glass and Entropy, *J. Non-Cryst. Sol.* **355** (2009) 581-758.

25. A. Takada, R.Conradt and P. Richet, *J. Non-Cryst. Sol.* **429** (2015) 33.

26. R.F. Tournier, *Physica B* **454**, (2014) 253.

27. R.F.Tournier and E. Beaugnon, *Sci. Technol.  Adv. Mater.* **10** (2009) 014501.

28. R.F. Tournier, *Metals.* **4** (2014) 359.

29. F.A. Lindemann, *Phys. Z.* **11** (1911) 609.

30. R.F. Tournier, *Physica B*  **392** (2007) 79.

31. T.R. Kirkpatrick, D. Thirumalai and P.G. Wolynes, Phys. Rev. A **40** (1989) 1045.

32. J. Souletie, *J. Phys. France* **51** (1990) 883**.**

33. L. Berthier, G.Biroli, J. Bouchaud, L. Cipelletti, D.E. Masri, D.L'Hôte, F. Ladieu and M. Pierno, *Science* **310** (2005) 1797.

34. C.A. Angell, in *Structural Glasses and Supercooled Liquids, edited by* P.G. Wolynes and V. Lubchenko, (Wiley and Sons, Hoboken, New Jersey, 2012). p. 237.

35. M.I. Ojovan and W.E. Lee, *J. Non-Cryst. Sol.* **356** (2010) 2534.

36. M.I. Ojovan, K.P. Travis, and R.J. Hand, *J. Phys. Condens. Matter.* **19** (2007) 415107.

37. J.F. Stanzione III, K.E. Strawhecker, and R.P. Wool, *J. Non-Cryst. Sol.* **357** (2011) 311.

38. R.P. Wool, *J. Polymer. Sci. B* **46** (2008) 2765.

39. R.P. Wool and A. Campanella, *J. Polymer. Phys. part B: Polym. Phys.* **47** (2009) 2578.

40. D.S. Sanditov, *J. Non-Cryst. Sol.* **385** (2014) 148.

41. R.F. Tournier, *Chem. Phys. Lett.* **641** (2015) 9.

42. K. Yamamoto, Y. Shibayama, and K. Shirahama, *J. Phys. Soc. Jpn*. **77** (2008) 013601

43. D.N. Bittner and E.D. Adams, *J. Low Temp. Phys.* **97** (1994) 519.

44. K.L. Kearns, K.R. Whitaker, M.D. Ediger, H. Huth and C. Schick, J. Chem. Phys. **133**, (2010) 014702.